\def\eqnum#1{\eqno (#1)}
\def\fnote#1{\footnote}

\documentstyle[12pt]{article}
\font\tenbm=cmmib10
\font\sevenbm=cmmib7
\font\fivebm=cmmib5
\newfam\bmfam
\textfont\bmfam=\tenbm \scriptfont\bmfam=\sevenbm
\scriptscriptfont\bmfam=\fivebm
{\count0=\number\bmfam \multiply\count0 by "100
\def\defbgreek#1#2#3{{\count1=\count0 \advance\count1 by "#2#3
\global\mathchardef#1=\count1 }}
\defbgreek\balpha  0B \defbgreek\brho       1A
\defbgreek\bbeta   0C \defbgreek\bsigma     1B
\defbgreek\bgamma  0D \defbgreek\btau       1C
\defbgreek\bdelta  0E \defbgreek\bupsilon   1D
\defbgreek\bepsilon0F \defbgreek\bphi       1E
\defbgreek\bzeta   10 \defbgreek\bchi       1F
\defbgreek\bmeta   11 \defbgreek\bpsi       20
\defbgreek\btheta  12 \defbgreek\bomega     21
\defbgreek\biota   13 \defbgreek\bvarepsilon22
\defbgreek\bkappa  14 \defbgreek\bvartheta  23
\defbgreek\blambda 15 \defbgreek\bvarpi     24
\defbgreek\bmu     16 \defbgreek\bvarrho    25
\defbgreek\bnu     17 \defbgreek\bvarsigma  26
\defbgreek\bxi     18 \defbgreek\bvarphi    27
\defbgreek\bpi     19}
\title{Interplay between Dynamic Systems Described by the Klein-Gordon
and Dirac Equations}
\author{Yuri A.Rylov}
\date{Institute for Problems in Mechanics, Russian Academy of Sciences,\\
101, Vernadskii Ave.,  Moscow, 117526, Russia.\\
e-mail: rylov@ipmnet.ru}

\begin{document}

\maketitle


\begin{abstract}
It is shown that in the two-dimensional space-time the dynamic system,
described by the free Klein-Gordon equation, turns to the dynamic system,
described by the free Dirac equation, provided the current and the
energy-momentum tensor are redefined in a proper way.
\end{abstract}

The purpose of the present paper is to compare dynamic properties of
two dynamic systems ${\cal S}_{{\rm KG}}$ and ${\cal S}_{{\rm D}}$
in the  two-dimensional space-time, where ${\cal S}_{{\rm KG}}$
is a dynamic system, described by the free Klein-Gordon equation and
${\cal S}_{{\rm D}}$ is the dynamic system, described by the free
Dirac equation.
Such a comparison of only dynamic properties is of interest in the light  of
the fact \cite{R98} that quantum effects can be explained in
terms of dynamics only (i.e. without a reference to quantum axiomatics).

Conventionally the term dynamic system (dynamics) means a mathematical
object, whose state is described by some dynamical variables, which may be
real- or complex-valued quantities. Any physical quantity is a function
of the state and can be expressed through the dynamic variables. Dynamic
variables evolve according to dynamic equations which determine
single-valuedly evolution of the state. Sometimes mathematical objects whose
dynamical variables are operator-valued, or matrix-valued quantities are
also considered as dynamic systems (so called quantum dynamics), although
in this case the physical quantities cannot be expressed only through
dynamic variables. For calculation of physical quantities one needs
to introduce additional quantity (the state vector). As a result any
physical quantity is expressed via dynamic variables and the state vector.

We shall use a more narrow definition. {\it Dynamic system is a set of dynamic
variables, fully describing the state of that system, dynamic equations
for them and expressions for the current $j^l $ and the energy-momentum tensor
$T^{kl}$}. Both dynamic equations and expressions for $j^l$ and $T^{kl}$ can
be obtained from the action functional. In other words, any dynamic system
is determined by the action functional which contains dynamic
variables as its arguments.

Let us discuss in more detail  the  constraint  that  dynamic  variables
must fully describe the state of the system. The meaning of  this
constraint can be illustrated using an example of a free quantum  particle
${\cal S}_{\rm S}$, described by the Schr\"odinger equation
$$
i\hbar {\partial \psi _{\rm S}\over\partial t}+{\hbar ^{2}\over 2m}\nabla ^{2}\psi _{\rm S}=0.
\eqnum{1}$$
\noindent Here $\psi _{\rm S}=\psi _{\rm S}(t,{\bf x})$ is  a  continuous
complex  dynamical  variable, which fully discribes the state of  the
quantum  particle ${\cal S}_{\rm S}$.
Eq.(1) is a dynamic equation for the dynamic variable $\psi _{\rm S}$.  If
$\psi _{\rm S}$ is known, the  state  of ${\cal S}_{\rm S}$  and  all  physical  quantities,
relating to ${\cal S}_{\rm S}$  are determined.
\par
The same quantum system ${\cal S}_{\rm S}$ can be described in the Heisenberg
representation in terms of  the  position  operator ${\bf \hat{q}}$  and  the
momentum operator {\bf \^{p}}. The operators ${\bf \hat{q}}$ and ${\bf \hat{p}}$  are  considered  as
quantum dynamic variables. These variables depend on time $t$  and
satisfy the dynamic equations
$$
{{\rm d}{\bf \hat{q}}\over {\rm d}t}={{\bf \hat{p}}\over m},\qquad {{\rm d}{\bf \hat{p}}\over {\rm d}t}=0
\eqnum{2}$$
\noindent and the commutation relations
$$
[q^{\alpha },p_{\beta }]_{-}=i\hbar \delta ^{\alpha }_{\beta },\qquad \alpha ,\beta =1,2,3
\eqnum{3}$$
\noindent As far as dynamic equations (2) remind formally Hamilton
equation for a classical particle, the operators ${\bf q}$ and ${\bf p}$
are considered usually as quantum dynamic variables. Quantum dynamic
variables ${\bf \hat{q}}(t), {\bf \hat{p}}(t)$  are  quite different  from
dynamic variables $\psi _{\rm S}(t,{\bf x})$ in the sense that ${\bf \hat{q}}$,
${\bf \hat{p}}$  cannot fully describe the state of ${\cal S}_{\rm S}$.
In the Heisenberg representation  the state of ${\cal S}_{\rm S}$ is
determined by the state vector $\psi _{\rm H}$. All  physical
quantities relating to ${\cal S}_{\rm S}$  are  determined  by  quantum  dynamic
variables ${\bf \hat{q}}, {\bf \hat{p}}$ taken together with the state vector
$\psi _{\rm H}$ (but  not by dynamic variables only as it was in the
Schr\"odinger representation (1)).
\par
As we have shown, our definition of  the  dynamic  system  (and  dynamic
properties)  disqualifies Heisenberg representation of a  quantum
dynamic system  on the basis of its inability to describe the system
state using only dynamic variables and the need for additional information
in  the state vector $\psi _{\rm H}$. The underlying reason  of this
deficiency is that the Heisenberg description (2), (3) make use of
quantum  axiomatics  (quantum  principles)  which in fact have a little
to do with the dynamics of a system.
\par
The Schr\"odinger description, including  dynamic  equation  of
the type (1) and expressions for $j^{l}$ and $T^{kl}$, is sufficient  for
explanation  and  calculation  of  all  real   quantum   effects
\cite{R98}, but it does  not  contain  a  concept  of  a  single
particle. To describe and to interpret the  Schr\"odinger  picture
in  terms  of  a  single  particle,  one  uses  the  statistical
propositions which permit to calculate average  values $\langle R
\rangle _{\psi }$  of
any physical quantity $R$ at  the  state  described  by  the  wave
function (state vector) $\psi $
$$
\langle R\rangle _{\psi }= A^{-1}\int \psi ^{*}\hat{R}\psi d{\bf x},
\qquad A=\int \psi ^{*}\psi d{\bf x},
\eqnum{4}$$
\noindent where $\hat{R}$ is some linear operator  associated
with  the  physical
quantity $R$. Statistical propositions (4) is  some  kind  of  a
probabilistic construction that permits to speak  about  quantum
phenomena in terms of  a  single  particle.  It  is  a  kind  of
interpretation  of  quantum  phenomena  in  terms  of  a  single
particle.  As any interpretation this probabilistic construction
contains some arbitrary elements which  are  not  essential  for
calculation of quantum effects. Such an  interpretation  is  not
unique \cite{R98}.
\par
Quantum principles (quantum axiomatics) can  be  derived  from
the statistical propositions (4) provided they take  place
for any quantities \cite{N32}. The statistical propositions
agree with the dynamic equation (1) in such a  way  that  any
result which can be obtained from expressions for $j^{l}$ and $T^{kl}$ can
be obtained also from the statistical propositions (4).
\par
Quantum  dynamic   equations   (2)   in   the   Heisenberg
representation cannot be  derived  from  only  dynamic  equation
(1). They can be obtained only as a result of  combination  of
dynamic equation (1) with the statistical propositions (4).
Thus, looking formally as  dynamic  equations,  equations  (2)
contain implicitly statistical  propositions  (4)  and,  hence,
quantum axiomatics. In particular, the dynamic  equations  (2)
contain essentially the concept  of  linear  operator,  that  is
characteristic for the statistical propositions (4),  but  not
for the dynamic equation (1).
\par
In this paper only dynamic properties of two dynamic systems
${\cal S}_{\rm KG}$ and ${\cal S}_{\rm D}$ are compared, because these  properties  seem  to  be
most important for calculation  of  quantum  effects.  In  other
words, the properties connected with  dynamics  (1)  (but  not
with the statistical propositions)  are  investigated.  In  this
relation our investigation differs from  the  well-known paper by
Foldy-Wouthuysen \cite{FW50}   which  investigates  dynamic
equations for ${\cal S}_{\rm D}$ in the Heisenberg  representation  and,  hence,
takes  into  account  both  dynamics  and  quantum   axiomatics,
contained in the Heisenberg  dynamic  equations  (concept  of  a
linear operator, commutation relations, etc.)

Note that the paper \cite{R98} is necessary only for a motivation of the
presented investigation which is self-sufficient and does not need a
reference to the paper \cite{R98} for its substantiation. Necessity of the
investigation of only dynamics can be explained as follows.

Results of experiments with a single quantum particle are irreproducible.
By definition it means that the single quantum particle is stochastic.
Although a result of
an experiment with a single stochastic particle is irreproducible,
distributions of results of similar experiments with many independent identical
stochastic particles are reproducible. Projecting many independent
identical stochastic particles ${\cal S}_{\rm st}$ in the same space-time
region, one obtains a cloud ${\cal E}[N,{\cal S}_{\rm st}]$ of $N$
independent identical particles moving randomly.
With the number $N$ of particles tending to $\infty$, this cloud
${\cal E}[\infty ,{\cal S}_{\rm st}]$ may be considered as a continuous medium,
or a fluid. Only dynamics of the fluid is important for calculation of
physical phenomena connected with the stochastic particle ${\cal S}_{\rm st}$
This statement is valid for any stochastic particle independently
of the nature of the stochasticity. For instance, it is valid both for
a quantum particle associated with the quantum (Madelung) fluid and for a
Brownian particle associated with the Brownian fluid. In any case
this fluid is a deterministic dynamic system, because experiments
with this fluid ${\cal E}[\infty ,{\cal S}_{\rm st}]$ are reproducible.
Besides any reproducible experiment with the stochastic particle can be
described in terms of the fluid ${\cal E}[\infty ,{\cal S}_{\rm st}]$
without a reference to any probabilistic construction. Such constructions
(probability density, or probability amplitude) are needed only for
interpretation of the fluid motion in terms of a single stochastic particle.
For instance, instead of the density $\rho$ of the Brownian fluid, it is
a common practice to speak about a probability density $\rho$ of the
Brownian particle position. In the case of the quantum (Madelung) fluid
the interpretation in terms of a single particle is not so simple.

Thus, a stochastic particle associates with some kind of a fluid, and
studying dynamics of this fluid, one investigates mean properties of the stochastic
particle. The Madelung fluid is nondissipative, whereas the Brownian fluid
is dissipative, and this is a reason of different behaviour of qunatum
particles and Brownian ones.
\par
Conventionally ${\cal S}_{{\rm KG}}$
associates with a spinless particle and a scalar wave function $\psi$,
whereas ${\cal S}_{{\rm D}}$ associates with a particle of spin $1/2$ and a
spinor wave function $\psi _{\rm D}$. Here only the two-dimensional space-time
is considered, where spin is unessential. According to the quantum axiomatics
the transformation properties of wave functions are connected
with some internal properties of the described particle. If one abstracts
from the quantum axiomatics, and considers ${\cal S}_{{\rm KG}}$  and
${\cal S}_{{\rm D}}$ simply as dynamic systems, one
discovers that by means of a change of variables
the two-component spinor $\psi _{\rm D}$ can be transformed
into the one-component scalar $\psi$ and some combination of derivatives
of $\psi$. Under this transformation the Dirac equation for $\psi _{{\rm D}}$
transforms to the Klein-Gordon equation for $\psi$,  but
$j^l$ and $T^{kl}$ for ${\cal S}_{{\rm KG}}$ and ${\cal S}_{{\rm D}}$
do not transform one into other.

It means that from the dynamic point of view the dynamic systems
${\cal S}_{{\rm KG}}$ and ${\cal S}_{{\rm D}}$ differ only by definitions
of $j^l$ and $T^{kl}$.
Redefining  $j^l$ and $T^{kl}$, one can turn  ${\cal S}_{{\rm D}}$ to
${\cal S}_{{\rm KG}}$ and vice versa.

All this looks rather unexpected, if, {\sl basing on the quantum axiomatics},
the transformation properties
of the wave function are considered as internal properties of the
described  particle, because it is difficult to believe that internal
properties of the particle can depend on a choice of dynamic variables.

It should stress in this connection that our results contradict by no
means to the conventional results concerning the Dirac spinor field
$\psi _{\rm D}$ and the Klein-Gordon scalar field $\psi$, because they
concern quite different mathematical objects. Our results concern
properties of dynamic systems ${\cal S}_{\rm D}$ and ${\cal S}_{\rm KG}$
taken in themselves, whereas the Dirac spinor field
$\psi _{\rm D}$ and the Klein-Gordon field $\psi$
are defined as mathematical objects satisfying
some dynamic equations and in addition the constraints of quantum axiomatics.
In particular, these constraints concern transformation properties of
the field $\psi _{\rm D}$ and $\psi$. The $\psi _{\rm D}$ transforms as
a spinor, and any transformation of $\psi _{\rm D}$ into a scalar is
forbidden. Such a transformation is considered as incompatible with the
quantum axiomatics. In the pure dynamics there are no such constraints.
Any change of dependent variables is possible.

In the two-dimensional space-time the dynamic system ${\cal S}_{{\rm KG}}$
 is described by the action
$$
{\cal A}_{{\rm KG}}[\psi ,\psi ^{*}]={1\over 2}\int (-m^{2}c^{2}\psi ^{*}\psi
+\hbar ^{2}\partial _{l}\psi ^{*}\partial ^{l}\psi )d^{2}x,\qquad
\partial _l\equiv{\partial\over\partial x^l},\qquad l=0,1
\eqnum{5}$$
\noindent where $\psi$ is a one-component complex variable,
$m$ is a mass of a particle, $c$ and $\hbar $  are  respectively  the
speed of the light and the Planck constant, and $g_{ik}$=diag$\{c^{2},-1\}$,
$g^{ik}$=diag$\{c^{-2},-1\}$ is the metric  tensor  in  the  two-dimensional
space-time. There is a summation  over  repeated  Latin  indices
0-1.
The dynamic equation has the form
$$
\lambda ^{2}\partial _{l}\partial ^{l}\psi +\psi =0,\qquad \lambda \equiv \hbar /mc
\eqnum{6}$$
\noindent The current $j^{l}$ and the canonical energy-momentum tensor $T^{i}_{k}$  are
defined by the relations
$$
j^{l}={i\hbar \over 2}(\psi ^{*}\partial ^{l}\psi -\psi
\partial ^{l}\psi ^{*}),\qquad l=0,1
\eqnum{7}$$
$$
T^{l}_{k}={\hbar ^{2}\over 2}(\psi ^{*l}\psi _{k}+\psi ^{*}_{k}\psi ^{l}-
\delta ^{l}_{k}\psi ^{*}_{i}\psi ^{i})+{m^{2}c^{2}\over 2}\psi ^{*}\psi \delta ^{l}_{k},\qquad l,k=0,1
\eqnum{8}$$
$$
\psi _{k}\equiv \partial _{k}\psi ,\qquad \psi ^{k}\equiv \partial ^{k}\psi \equiv g^{kj}\partial _{j}\psi ,
\eqnum{9}$$
The current $j^{l}$  and  the  energy-momentum  tensor $T^{l}_{k}$  are
attributes of the dynamic system ${\cal S}_{{\rm KG}}$, because they  are  sources
of the electromagnetic field  and  of  the  gravitational  field
respectively. It means that the expression (7) for the current
determines interaction of ${\cal S}_{{\rm KG}}$ with the external  electromagnetic
field by means of additional term $-ec^{-1}A_{k}j^{k}$ in the Lagrangian  of
the action (5). The canonical energy-momentum tensor (8)  is
symmetric  and  coincides  with  that  derived  by  means  of  a
variation with respect to metric tensor $g_{ik}$.

The dynamic system ${\cal S}_{\rm D}$  is described by the action
$$
{\cal A}_{\rm D}[\psi _{\rm D},\psi ^{*}_{\rm D}]=\int (-mc\bar{\psi }_{\rm D}\psi _{\rm D}+{i\over 2}\hbar \bar{\psi }_{\rm D}\gamma ^{l}\partial _{l}\psi _{\rm D}-{i\over 2}\hbar \partial _{l}\bar{\psi }_{\rm D}\gamma ^{l}\psi _{\rm D})d^{2}x
\eqnum{10}$$
The  current $j_{\rm D}^k$ and   the   canonical
energy-momentum tensor $T^l_{{\rm D}k}$ have the form
$$
j^{l}_{{\rm D}}=\bar{\psi }_{{\rm D}}\gamma ^{l}\psi _{{\rm D}},\qquad l=0,1,
\eqnum{11}$$
$$
T^{l}_{{\rm D} k}={i\hbar \over 2}(\bar{\psi }_{{\rm D}}\gamma ^{l}\partial _{k}\psi _{{\rm D}}-\partial _{k}\bar{\psi }_{{\rm D}}\gamma ^{l}\psi _{{\rm D}}),\qquad l,k=0,1
\eqnum{12}$$
\noindent where
$$
\psi _{\rm D}=(^{\psi _{+}}_{\psi _{-}}),\qquad \bar{\psi }_{{\rm D}}=\psi _{\rm D}^*
c\gamma ^{0},\qquad \psi ^{*}=(\psi ^{*}_{+},\psi ^{*}_{-})
\eqnum{13}$$
$$
\gamma ^{0}=c^{-1}
\left(
\begin{array}{cc}
0 & 1\\
1 & 0
\end{array}\right) ,
\qquad \gamma ^{1}=\left(
\begin{array}{cc}
0 & 1\\
-1 & 0
\end{array}\right) ,\qquad
\gamma _{0}=g_{00}\gamma ^{0}=c
\left(
\begin{array}{cc}
0 & 1\\
1 & 0
\end{array}\right) ,
\eqnum{14}$$
\par
Representation (14) of $\gamma $-matrices is chosen in such a way that
the  pseudo-scalar  matrix $c\gamma ^{0}\gamma ^{1}$  is  diagonal,  and  the  wave
functions $\psi _{{\rm D}}=(^{\psi _{+}}_{0}),\quad \psi _{\rm D}=(^{0}_{\psi_-})$
are  its  eigenfunctions  for  any
choice of $\psi _{+}, \psi _{-}$.  In  this case in virtue of Eq.(13)
the Dirac equation
$$
i\hbar \gamma ^{l}\partial _{l}\psi _{{\rm D}}-mc\psi _{{\rm D}}=0
\eqnum{15}$$
\noindent takes the form
$$
\psi _{+}=i\lambda \partial _{+}\psi _{-},\qquad \psi _{-}=i\lambda
\partial _{-}\psi _{+},
\eqnum{16}$$
$$
\lambda \equiv \hbar /mc,\qquad \partial _{\pm }\equiv c^{-1}\partial _{0}
\pm \partial _{1}
\eqnum{17}$$
\noindent It follows from Eq.(16) that both wave functions $\psi _{\pm }$
satisfy  the free Klein-Gordon equation
$$
\lambda ^2\partial _l\partial ^l\psi _{\pm }+\psi _{\pm}=0
\eqno (18)$$
\par
Let us introduce the two-component differential operator
$$
{\cal L}(w,\partial ,\lambda )=\left(
\begin{array}{c}
\sqrt{w_+}+i\lambda \sqrt{w_-}\partial _{+}\\
\sqrt{w_-}+i\lambda \sqrt{w_+}\partial _{-}
\end{array}\right).
\eqnum{19}$$

\noindent where $w_l=(w_0,w_1)$ is a constant timelike or null vector and
$$
w_{+}=c^{-1}w_{0}+w_{1},\qquad w_{-}=c^{-1}w_{0}-w_{1}
$$
Under the continuous Lorentz transformation
$$
x^{0}\to  \tilde{x}^{0}=x^{0}\cosh \chi +c^{-1}x^{1}\sinh \chi
$$
$$
x^{1}\to  \tilde{x}^{1}=x^{1}\cosh \chi +cx^{0}\sinh \chi
\eqnum{20}$$
\noindent the components $w_{\pm }$ and $\partial _{\pm }$ transforms as follows
$$
w_{+}\to \tilde{w}_{+}=e^{\chi }w_{+},\qquad w_{-}\to\tilde{w}_{-}=
e^{-\chi }w_{-}
$$
$$
\partial _{+}\to \tilde{\partial }_{+}=e^{\chi }\partial _{+},
\qquad \partial _{-}\to\tilde{\partial }_{-}=
e^{-\chi }\partial _{-}
\eqnum{21}$$
\noindent According to Eqs. (20), (21) the differential operator
${\cal L}$
transforms as follows
$$
{\cal L}(w,\partial ,\lambda )\to  {\cal L}(\tilde{w},\tilde{\partial },
\lambda )=\left(
\begin{array}{c}
\sqrt{\tilde{w}_+}+i\lambda \sqrt{\tilde{w}_-}\tilde{\partial }_{+}\\
\sqrt{\tilde{w}_-}+i\lambda \sqrt{\tilde{w}_+}\tilde{\partial }_{-}
\end{array}\right)=
$$
$$=\left(
\begin{array}{c}
 e^{\chi /2}(\sqrt{w_+}+i\lambda \sqrt{w_-}\partial _{+})\\
 e^{-\chi /2}(\sqrt{w_-}+i\lambda \sqrt{w_+}\partial _{-})
\end{array}\right)=e^{-c\gamma ^{0}\gamma ^{1}\chi /2}{\cal L}(w,
\partial ,\lambda )
\eqnum{22}$$
\noindent Under space reflections
$$
x^0\to \tilde{x}^{0}=x^{0},\qquad x^{1}\to \tilde{x}^{1}=-x^{1}
\eqnum{23}$$
one has
$$
w_{+} \to \tilde{w} _+=w_{-},\qquad w_{-}\to\tilde{w}_-=w_{+},
$$
$$
\partial _{+} \to \tilde{\partial }_+=\partial _{-},
\qquad \partial _{-}\to\tilde{\partial }_-=\partial _{+},
\eqno (24)$$
$$
{\cal L}(w,\partial ,\lambda )\to {\cal L}(\tilde{w},\tilde{\partial },
\lambda )=c\gamma ^{0}{\cal L}(w,\partial ,\lambda )
\eqnum{25}$$

Under time reflections
$$
x^{0}\to\tilde{x}^{0}=-x^{0},\qquad x^{1}\to\tilde{x}^{1}=x^{1}
\eqnum{26}$$
one can write
$$
w_{+}\to\tilde{w}_+=e^{i\pi }w_{-},\qquad w_{-}\to\tilde{w}_-=e^{-i\pi }w_{+},
$$
$$
\partial _{+}\to\tilde{\partial }_+=e^{i\pi }\partial _{-},
\qquad \partial _{-}\to\tilde{\partial }_-=e^{-i\pi }\partial _{+},
\eqnum{27}$$
$$
{\cal L}(w,\partial ,\lambda )\to {\cal L}(\tilde{w},\tilde{\partial },
\lambda )=e^{i\pi /2}\gamma ^{1}{\cal L}(w,\partial ,\lambda )
\eqnum{28}$$
\noindent It means that the differential operator
${\cal L}(w,\partial ,\lambda )$
transforms
as a spinor under all transformations of the Lorentz group.
\par
Let us form the two-component quantity
$$
\psi _{{\rm D}}=(^{\psi _+}_{\psi _-})={\cal L}(w,\partial ,\lambda )\psi
\eqnum{29}$$
\noindent If $\psi $ is a scalar, satisfying the KG equation (6), then $\psi _{{\rm D}}$ is a
spinor, satisfying the Dirac equation (15) for any choice of the
timelike constant vector $w_l=(w_{0},w_{1})$. Vice versa, if the spinor $\psi _{{\rm D}}$
satisfies the Dirac equation (15), then the scalar $\psi $ defined by
Eq.(29) satisfies the KG equation (6) for any choice of the timelike
vector $w$.

The transformation reciprocal to the transformation  (29)
can be presented in the explicit form
$$
\psi ={1\over 2}\hat{Q}(\sqrt{w_+}\psi _{+}+\sqrt{w_-}\psi _-)
\eqno (30)$$
$$
c^{-1}\partial _{0}\psi =-{i(1-c^{-1}w_{0}\hat{Q})\over 2\lambda
\sqrt{w_-}}\psi _{+}-{i(1+c^{-1}w_{0}\hat{Q})\over 2\lambda \sqrt{w_+}}
\psi _{-}
\eqno (31)$$
\noindent where $\hat{Q}=(w_{1}+i\lambda w\partial _{1})^{-1}$ is
the operator reciprocal to the operator
$w_{1}+i\lambda w\partial _{1}$
$$
\hat{Q}\psi (t,x)=(2\pi \hbar )^{-1}\int\limits^{\infty }_{-\infty }
\int\limits^{\infty }_{-\infty }
\exp[i\hbar ^{-1}k(x-x')](w_{1}-{kw\over mc})^{-1}\psi (t,x')dkdx'
\eqno (32)$$
$$
w\equiv \sqrt{\mathstrut w_+w_-}=\sqrt{\mathstrut w_kw^{k}}
$$

The state of ${\cal S}_{{\rm KG}}$ is described by $\psi$ and $\partial _0\psi$
given at some moment of time. The relations (30), (31) express these
quantities via components $\psi _+$ and $\psi _-$ of the spinor $\psi _{\rm D}$.
\par
Substituting relation (29) into Eq.(11), one obtains
$$
j^{0}_{{\rm D}}=c^{-1}(\psi ^{*}_{+}\psi _{+}+\psi ^{*}_{-}\psi _{-})=
{4\over m^{2}c^{2}}(T^{0}_{l}w^{l}+wmcj^0)\hbox{sgn}(w_0),
\qquad w\equiv\sqrt{w_lw^l}
\eqnum{33}$$
$$
j^{1}_{{\rm D}}=-\psi ^{*}_{+}\psi _{+}+\psi ^{*}_{-}\psi _{-}=
{4\over m^{2}c^{2}}(T^{1}_{l}w^l+wmcj^1)\hbox{sgn}(w_0),
$$
It means that  the  current $j_{{\rm D}}$  associated  with  the  Dirac
dynamic system is constructed of the components of the current $j^l$ and
the energy-momentum
tensor $T^{l}_{k}$, associated with the Klein-Gordon dynamic system.
\par
Using Eq.(12), (14), one can present components of $T^{l}_{{\rm D} k}$ in the form
$$
T^{0}_{{\rm D} l}=c^{-1}(j_{(+)l}+j_{(-)l}),\qquad T^{1}_{{\rm D} l}=j_{(-)l}-j_{(+)l},
\qquad l=0,1
\eqnum{34}$$
\noindent where $j_{(\pm )}$ are quantities of the type of the current (7)
$$
j_{(+)l}={i\hbar \over 2}(\psi ^{*}_{+}\partial _{l}\psi _{+}-\psi _{+}
\partial _{l}\psi ^{*}_{+}),
\qquad j_{(-)l}={i\hbar \over 2}(\psi ^{*}_{-}\partial _{l}\psi _{-}-\psi _{-}\partial _{l}\psi ^{*}_{-})
\eqnum{35}$$
Let us note that the canonical energy-momentum  tensor $T^{l}_{{\rm D} k}$
is symmetric in virtue of dynamic equations, although it is  not
symmetric identically. Indeed, using relations (16) between  the
components $\psi _{+}$ and $\psi _{-}$ and Eq.(18), one can show that
$$
j_{(+)1}+j_{(-)1}=c^{-1}(j_{(+)0}-j_{(-)0}).
\eqnum{36}$$
\noindent It follows from Eqs.(34) and (36)  that
$$
c^{2}T^{0}_{{\rm D} 1}=-T^{1}_{{\rm D} 0},\qquad T^{01}_{{\rm D}}=T^{10}_{{\rm D}}
\eqnum{37}$$

To present the energy-momentum tensor $T^l_{{\rm D}k}$ in terms of $\psi $
in an explicit form, one introduces expressions
$$
{\cal T}^l_k(\chi ^*,\psi )=
{m^{2}c^{2}\over 2}\chi ^{*}\psi \delta ^{l}_{k}
+{\hbar ^{2}\over 2}(\chi ^{*l}\psi _{k}+\chi ^{*}_{k}\psi ^{l}-
\delta ^{l}_{k}\chi ^{*}_{i}\psi ^{i}),\qquad l,k=0,1
\eqno (38)$$
$$
\chi _l\equiv\partial _l\chi, \qquad
\psi _l\equiv\partial _l\psi
$$
$$
{\cal J}^l(\chi ^*,\psi)
={i\hbar \over 2}(\chi ^{*}\partial ^{l}
\psi -\psi \partial ^{l}\chi ^{*}),\qquad l=0,1
\eqno (39)$$
\noindent where $\chi ^*$ and $\psi$ are formal arguments of the functions
${\cal T}^k_l$ and ${\cal J}^l$.
According to Eqs. (11), (12) the energy-momentum tensor $T^l_{{\rm D}k}$
is a result of the operator $i\hbar\partial _k$ action on the wave
function $\psi _{\rm D}$ in the expression (11) for $j^l_{\rm D}$. Then according
to Eq.(8) and Eqs.(38), (39) one obtains
$$
T^l_{{\rm D}k}
={2i\lambda\over mc}\hbox{sgn}(w_0)\left\{
{\cal T}^l_s(\psi ^*,\psi _k)w^s-{\cal T}^l_s(\psi ,\psi ^*_k)w^s
\right. +
$$
$$
\left. +wmc[{\cal J}^l(\psi ^*,\psi _k)-{\cal J}^l(\psi ,\psi ^*_k)]\right\}
,\qquad\psi _k\equiv\partial _k\psi
\eqno (40)$$
These expressions are relativistically covariant with respect to scalars
$\psi , \psi ^*$ and the vector $w_l$.

Thus, if $\psi $ satisfies the KG equation (6),
the $\psi $ describes the dynamic system ${\cal S}_{{\rm KG}}$, provided
the current $j^{k}$ and the energy-momentum tensor $T^{l}_{k}$ are
defined by Eqs. (7) and (8) respectively.
\par
If $\psi $ satisfies the same KG equation (6), but  the  current
$j^{k}$ and the energy-momentum tensor $T^{l}_{k}$  are  determined  by  Eqs.
(33), and (38)--(40)  respectively, the $\psi$ describes the dynamic
system ${\cal S}_{\rm D}$.
\par
In other words, dynamic systems ${\cal S}_{{\rm KG}}$ and
${\cal S}_{{\rm D}}$ differ
by definition of the current and the energy-momentum tensor (but
not  by  their  dynamic   equations).   Interaction   with   the
electromagnetic field is determined by the form of  the  current.
It is different for ${\cal S}_{{\rm KG}}$ and ${\cal S}_{{\rm D}}$.

Description of the dynamic system ${\cal S}_{\rm D}$ in terms of the scalar
$\psi$ contains a timelike or null vector  $w_l$. Description in terms of
the spinor $\psi _{\rm D}$ does not contain this vector $w_l$, because it is
"hidden inside $\gamma ^l$". A similar situation arises in the case of
the conventional 4-dimensional Dirac equation \cite{R995}. In this case
a constant timelike 4-vector arises, if the Dirac system is described
in terms of scalar-tensor variables (not in terms of spinors).

The transformation properties of a field are usually considered to be
its important characteristic. In  this  connection  the
statement that the dynamic system ${\cal S}_{{\rm D}}$, described usually in terms
of a spinor field $\psi _{{\rm D}}$, can be also described in terms of a scalar
field $\psi $  looks  rather  unexpectedly.  To  clear  the  situation,
following Anderson \cite{A67}, let us introduce the concept of an
absolute object. By the definition  the  absolute  object  is  a
quantity (or quantities) which is the same for all solutions  of
dynamic equations.
The  absolute  objects  arise,  if  the  dynamic  equations  and
expressions for $j^{k}$ an $T^{kl}$ are written  in  the  relativistically
covariant form. In  many  cases  the  absolute  objects  may  be
substituted by numbers, or some definite functions. But in  this
case the form of dynamic equations and expressions for $j^{k}$ an $T^{kl}$
stops to be covariant.
The total symmetry of the dynamic system is determined
by the symmetry of the absolute objects.
For instance, the description of ${\cal S}_{\rm KG}$, by the
action (1) [by the dynamic equation (6), the current (7) and
the energy-momentum tensor (8)] contains the metric tensor
$g_{lk}$=diag$\{ c^{2},-1\} $ which is an absolute object, because
it is the same for all solutions of the dynamic equation (6).
Components of $g^{kl}$ are invariant with respect to Lorentz trnsformations.
Symmetry of
${\cal S}_{\rm KG}$
is determined by the Lorentz group, and ${\cal S}_{\rm KG}$ is a relativistic
dynamic system.

A description of the dynamic system ${\cal S}_{{\rm D}}$, described
by the action (10) [by the dynamic equation (15), the current (11)
and the energy-momentum tensor (12)] contains the absolute objects
$\gamma ^l,\quad l=0,1$.
Such a description of ${\cal S}_{{\rm D}}$  does not contain the metric tensor
directly. The last arises as a derivative absolute object via the relations
$$
\gamma ^{l}\gamma ^{k}+\gamma ^{k}\gamma ^{l}=2g^{kl},\qquad k,l=0,1.
\eqnum{41}$$
\noindent The same dynamic system ${\cal S}_{{\rm D}}$, described by the
dynamic equation  (6), the current (33), (7), (8) and the
energy-momentum tensor (40), contains two absolute objects: the metric
tensor $g^{ik}$ and the constant timelike vector $w_l$. Apparently, this
fact would be interpreted in the sense that the change of variables
(29), (19) replaces the absolute objects $\gamma ^l,\quad l=0,1$
by the absolute objects $g^{ik}$, $w_l$. It correlates with the paper
\cite{R995}, where the same result was obtained for the Dirac dynamic system
in the $4D$ space-time. Although in the paper \cite{R995} another change of variables
was used, but that change of variables removed $\gamma$-matrices and lead also
to appearance of a new absolute object: a constant timelike vector $f^k$
instead of $\gamma $-matrices.

The Lorentz group is a symmetry group of the metric tensor $g^{kl}$,
but it is not a symmetry group of the constant vector $w_l$. It means that
the dynamic system
${\cal S}_{\rm D}$
described in terms  of the scalar $\psi$ is non-relativistic. It is
connected with the fact that the timelike vector $w_l$ means a preferred
direction in the space-time, or a preferred coordinate system, or a split
of the space-time into the space and the time.

At the same time the ${\cal S}_{\rm D}$ described in terms of the spinor
$\psi _{\rm D}$ is considered conventionally as a relativistic dynamic
system, because the absolute objects $\gamma ^l$ are considered as
invariants with respect to the Lorentz group. Does it mean that one can
convert a relativisitic dynamic system to non-relativistic one by means of
a change of variables? Such a possibility seems rather doubtful.
As far as the ${\cal S}_{{\rm D}}$,  described in terms of $\psi _{{\rm D}}$,
contains  only $\gamma ^{l},\quad l=0,1$ as original absolute objects, a
decision  on  relativistic
character of ${\cal S}_{{\rm D}}$ depends completely on the symmetry group of  the
$\gamma $-matrices $\gamma ^{l}$, or on whether or not $\gamma ^{l}$  are  scalars
under  the
transformations of the Lorentz group. $\gamma ^{l},\quad l=0,1$ are not objects
of the space-time, and transformation property of $\gamma ^{l}$ under the
Lorentz group is a rather subtle question.
\par

There are two approaches to the Dirac equation.
In the first approach \cite{S30} the wave function $\psi _{\rm D}$
is considered as a scalar function defined on the field of
Clifford numbers $\gamma ^l$,
$$
\psi _{\rm D}=\psi _{\rm D}(x,\gamma )\Gamma,\qquad
\overline{\psi }_{\rm D}=\Gamma\overline{\psi }_{\rm D}(x,\gamma ),
\eqno (42)$$
where $\Gamma$ is a constant nilpotent factor which
has the property $\Gamma f(\gamma )\Gamma =a\Gamma $.
Here $f(\gamma )$ is arbitrary function of $\gamma ^{l}$ and $a$ is
a complex number depending on the form of the function $f$.
Within such an approach under the Lorentz  transformations $\bar{\psi }_{{\rm D}}$,
$\psi _{{\rm D}}$
transform as scalars and $\gamma ^{l}$ transform as components of a vector.
In  this
case the symmetry group of $\gamma ^{l}$ is  a  subgroup   of  the  Lorentz
group, and ${\cal S}_{{\rm D}}$  is  non-relativistic  dynamic  system.  Then  the
matrix vector $\gamma ^{l}$  describes  some  preferred  direction  in  the
space-time.

In the second (conventional) approach the  $\psi _{\rm D}$ is
considered as a spinor, and the $\gamma ^l,\quad l=0,1$ are scalars with
respect to the transformations of the Lorentz group.
Analyzing the two approaches, Sommerfeld \cite{S} considered the first
approach as more reasonable.
In the second case the analysis is rather  difficult  due  to  of
non-standard transformations of $\gamma ^{l}$ and $\psi _{\rm D}$ under linear coordinate
transformations $T$. Indeed, the transformation $T$ for the vector $j^{l}$
$$
\tilde{\overline{\psi }}_{\rm D}\tilde{\gamma }^{l}\tilde{\psi }_{\rm D}={\partial \tilde{x}^{l}
\over \partial x^{s}}\overline{\psi }_{\rm D}\gamma ^{s}\psi _{\rm D},
\eqnum{43}$$
\noindent where quantities marked by tilde {\char"7E} mean quantities  in
the  transformed  coordinate system, can be written by two ways
$$
(1):\qquad\tilde{\psi }_{\rm D}=\psi _{\rm D},\qquad\tilde{\overline{\psi }}
_{\rm D}=\overline{\psi }_{\rm D},\qquad \tilde{\gamma }^{l}={\partial \tilde{x}^{l}\over \partial x^{s}}\gamma ^{s}
\eqnum{44}$$
$$
(2):\qquad \tilde{\gamma }^{l}=\gamma ^{l},\qquad \tilde{\psi }_{\rm D}=
S(\gamma ,T)\psi _{\rm D},\qquad \tilde{\overline{\psi }}_{\rm D}=\overline{\psi }
_{\rm D}S^{-1}(\gamma ,T),
\eqnum{45}$$
$$
S^{*}(\gamma ,T)\gamma ^{0}=\gamma ^{0}S^{-1}(\gamma ,T)
$$
\noindent The relation (44) corresponds to the first  approach  and  the
relation (45) to the second one. Both ways (44)  and  (45)
lead to the same result, provided
$$
S^{-1}(\gamma ,T)\gamma ^{l}S(\gamma ,T)={\partial \tilde{x}^{l}\over \partial x^{s}}\gamma ^{s}
\eqnum{46}$$
\noindent The second way (45) has two defects. First, the transformation
law  of $\psi _{\rm D}$  depends  on $\gamma $,  i.e.   under   linear   coordinate
transformation $T$ components of $\psi _{\rm D}$ transform through $\psi _{\rm D}$  and $\gamma ^{l}$,
but  not  only  through $\psi _{\rm D}$.  Second,  the  relation  (46)  is
compatible with Eq.(41) only  under  transformations $T$  between
orthogonal coordinate systems, when components $g^{lk}=\{c^{-2},-1\}$  of
the metric tensor are invariant. In other words, at  the  second
approach the relation (41) is not covariant with respect to arbitrary
linear transformations of coordinates. In this  case  the
symmetry group of  the  dynamic  system  does  not  coincide  in
general with the symmetry group of absolute objects. Due to  the
two  defects  of   the   conventional   (second)   approach   an
investigation becomes  rather  difficult. Within the  second
approach it is rather difficult to discover, where  "the  vector
$w^{l}$ is hidden". Within the first approach, when $\gamma ^{l}$
transforms  only through $\gamma ^{l}$, and $\psi _{\rm D}$
transforms only through $\psi _{\rm D}$, it is clear  that
the vector $w_{l}$ arises from the vector matrices $\gamma ^{l}$.
\par
Thus, for describing ${\cal S}_{{\rm D}}$  one may use a constant
vector $w_{l}$ and  a  scalar
field $\psi $ instead of the spinor $\psi _{\rm D}$. The fact that $\psi _{{\rm D}}$ defined by the
relations (19), (29) is a spinor follows directly  from  the
fact that $\psi $ is a scalar field and $w_{l}$ is a  constant  vector.  So
defined $\psi _{{\rm D}}$ is  a  spinor  independently  of  whether  or  not $\psi $
satisfies the KG equation (6). But the fact that $\psi _{{\rm D}}$ defined by
Eqs.(19), (29) satisfies  the  Dirac  equation  (15)  is  a
corollary of the fact that $\psi $ satisfies Eq.(6).
\par
Let us note that Eqs. (29), (19) can be written in  the
form which does not depend on reperesentation of $\gamma $-matrices,
defined by Eq.(41). Eq.(29) can be written in the form
$$
\psi _{{\rm D}}=[\Pi _-(\sqrt{w_+}+i\lambda \sqrt{w_-}\partial _{+})+
\Pi _+(\sqrt{w_-}+i\lambda \sqrt{w_+}\partial _{-})]
\left(
\begin{array}{c}
\psi\\
\psi
\end{array}\right)
\eqnum{47}$$
$$
\Pi _{\pm }={1\over 2}(1\pm c\gamma ^{0}\gamma ^{1})
\eqnum{48}$$
Thus, the spinor field in  the  two-dimensional  space-time
can be considered as a combination of a constant timelike (or null)
vector and a scalar field that associates with the Kramers
transformation \cite{K57}.
\par
Dynamic systems ${\cal S}_{{\rm D}}$ and ${\cal S}_{{\rm KG}}$ differ  only  by  the
definition of the current $j^{l}$ and the energy-momentum tensor $T^{l}_{k}$
which  are  sources  of  the  electromagnetic  field   and   the
gravitational field respectively. Both dynamic  systems  can  be
described either by the spinor field $\psi _{{\rm D}}$, or by the scalar  field
$\psi $. If one describes ${\cal S}_{{\rm D}}$  in terms of  the  scalar  field $\psi $,  the
choice of the constant vector $w_{l}$ is arbitrary that leads to some
arbitrariness of expressions for $j^{l}$ and $T^{l}_{k}$ in terms of $\psi $.
Description of ${\cal S}_{{\rm KG}}$ in  terms  of  the  spinor  field $\psi _{{\rm D}}$
is formally possible also. But it contains too many absolute objects:
$g_{kl}$, $\gamma ^l$, and $w_l$.
\par
Let us consider the case, when $\psi =ae^\kappa$, $a=$const is a real
wave function. Then $j^{k}\equiv 0,\quad k=0,1,$  and
the current vanishes in ${\cal S}_{{\rm KG}}$.
In this case according to Eqs.(19), (33)-(35) the  dynamic
system ${\cal S}_{{\rm D}}$ is described by the following quantities:
$$
j^{l}_{{\rm D}}=2ae^{2\kappa }[w^l+\lambda ^{2}(2\kappa ^{l}
\kappa _{s}w^s-w^l\kappa _s\kappa ^s)]\hbox{sgn}(w_0)
\eqnum{49}$$
$$
T^{l}_{{\rm D}k}=0,\qquad l,k=0,1
\eqnum{50}$$

Thus, the real $\psi $ describes a vanishing current in the dynamic
system ${\cal S}_{{\rm KG}}$ and a vanishing energy-momentum tensor in the dynamic
system ${\cal S}_{{\rm D}}$. It means essentially that one interchanges  roles  of
the current and energy-momentum in dynamic systems ${\cal S}_{{\rm KG}}$  and ${\cal S}_{{\rm D}}$.
Indeed, in ${\cal S}_{{\rm KG}}$  the  energy  density $T^{0}_{0}\ge 0,$  but
the  particle
density $j^{0}$ can be both positive and negative. Vice versa, in ${\cal S}_{{\rm D}}$
the particle density $j_{\rm D}^{0}\ge 0,$ but the energy density  can  be  both
positive and negative. The situation, when the energy density is
non-negative, but the particle density can be negative,  seems  more
preferable from physical viewpoint, than the situation, when the
particle density is non-negative,  but  the  energy  density  is
negative. The negative $j^{0}$ can be interpreted  as  a  density  of
antiparticles, but it is very difficult to interpret the  states
with negative energy of free particles. Such  states  should  be
only removed. (A possibility of the second quantization is not considered,
because it uses the quantum axiomatics essentially, whereas  here only
dynamic properties are considered).
\par
Thus, the dynamic system ${\cal S}_{\rm D}$ has two defects: (1) states with a
negative energy, (2) incompatibility of the definition of the current and the
energy-momentum tensor with the relativity principle.
The KG dynamic system ${\cal S}_{{\rm KG}}$ seems to be preferable,  than
${\cal S}_{{\rm D}}$, provided one does not appeal to the quantum axiomatics.
\par
To investigate  the  dynamic  systems ${\cal S}_{{\rm KG}}$   and ${\cal S}_{{\rm D}}$  in  the
non-relativistic approximation, let us make the transformation
$$
\psi =m^{-1/2}\exp (-i\hbar ^{-1}mc^{2}t)\Psi
\eqnum{51}$$
\noindent Then the action (5) turns to the action
$$
{\cal A}_{{\rm KG}}[\Psi ,\Psi ^{*}]=\int \{{i\hbar \over 2}(\Psi ^{*}\partial _{0}\Psi -\partial _{0}\Psi ^{*}\Psi )-{\hbar ^{2}\over 2m}\partial _{1}\Psi ^{*}\partial _{1}\Psi +{\hbar ^{2}\over 2m}\partial _{0}\Psi ^{*}\partial _{0}\Psi \}d^{2}x
\eqnum{52}$$
\noindent If characteristic frequencies $\omega $ of the wave function $\Psi $ are small
with respect to $mc^{2}/\hbar $, the last term  in  the  action  (52)  is
small with respect to others. Formally it is of the order of $c^{-2}$.
It  can  be  neglected.  Then  one
obtains the action for the non-relativistic Schr\"odinger equation.
\par
Substituting Eq.(51) into Eqs.(8), (9), one obtains the
following expressions for the current $j^{k}$ and the energy-momentum
$T^{k}_{l}$
$$
j^{0}=\Psi ^{*}\Psi +{i\hbar \over 2mc^{2}}(\Psi ^{*}\partial _{0}\Psi -\partial _{0}\Psi ^{*}\Psi )
\eqnum{53}$$
$$
j^{1}=-{i\hbar \over 2m}(\Psi ^{*}\partial _{1}\Psi -\partial _{1}\Psi ^{*}\Psi )
\eqnum{54}$$
$$
T^{0}_{0}=mc^{2}\Psi ^{*}\Psi +{\hbar ^{2}\over m}\Psi ^{*}_{1}\Psi _{1}+{\hbar ^{2}\over 4m}\partial _{k}\partial ^{k}(\Psi ^{*}\Psi )
\eqnum{55}$$
$$
T^{0}_{1}=mj_{1}+m{\lambda ^{2}\over 2}(\Psi ^{*}_{1}\Psi _{0}+\Psi ^{*}_{0}\Psi _{1})
\eqnum{56}$$
$$
T_{0}^{1}=-mc^{2}j_{1}+mc^{2}{\lambda ^{2}\over 2}(\Psi ^{*}_{1}\Psi _{0}+
\Psi ^{*}_{0}\Psi _{1})
\eqno (57)$$
$$
T^{1}_{1}=-{\hbar ^{2}\over m}\Psi ^{*}_{1}\Psi _{1}-{\hbar ^{2}\over 4m}
\partial _{k}\partial ^{k}(\Psi ^*\Psi )
\eqnum{58}$$
\noindent where one uses for brevity designations $\Psi _{k}\equiv
\partial _{k}\Psi ,\quad k=0,1.$  All
relations (52)-(58) are exact  relativistic  expressions.  The
last terms of Eqs.(52), (53),  (55), (56), (58)  have  either  the  order
$O(c^{-2})$, or have the form of a divergence. They can be  neglected
in the non-relativistic approximation. Eq.(56) can be written in the
approximate form
$$
T^{0}_{0}=mc^{2}j^0 +{\hbar ^{2}\over 2m}\Psi ^{*}_{1}\Psi _{1}+O(c^{-2}),
\eqnum{59}$$

\noindent where $j^0$ is determined by Eq.(53) to within $c^{-2}$.
\par
Now let us calculate in terms  of $\Psi $  the  current  and  the
energy-momentum tensor of  the  system ${\cal S}_{{\rm D}}$.  One  obtains  after
calculations
$$
j^{0}_{{\rm D}}=j^{0}+{i\hbar \over 2mc}(\Psi ^{*}\partial _{1}\Psi -
\partial _{1}\Psi ^{*}\Psi )+O(c^{-2})
\eqnum{60}$$
$$
j^{1}_{{\rm D}}=j^{1}
-{i\hbar \over 2mc}(\Psi ^{*}\partial _{0}\Psi -
\partial _{0}\Psi ^{*}\Psi )+O(c^{-2})
\eqnum{61}$$
\noindent where $j^{k}$ is determined by relations (53), (54). The  difference
between $j_{{\rm D}}$ and $j$ is  of  the  order $c^{-1}$  and  vanishes  in  the
non-relativistic approximation, when $c\rightarrow \infty $.


\end{document}